# Polarized beam operation of the Hybrid Spectrometer at the pulsed Spallation Neutron Source

I. Zaliznyak[1], V. Ghosh[2], S. M. Shapiro[1,2], L. Passell[2]

[1]Department of Physics, Brookhaven National Laboratory, 11973-5000, Upton, NY, USA

[2]Center for Neutron Science, Brookhaven National Laboratory, 11973-5000, Upton, NY, USA

**Abstract**

The concept of a neutron Hybrid Spectrometer (HYSPEC) combines the time-of-flight spectroscopy with the focusing Bragg optics and incorporates a polarized beam option. Here we describe the polarization analysis scheme proposed for HYSPEC and quantify its performance via the Monte-Carlo simulations. We find that the broadband supermirror-bender transmission polarizers provide reasonably good polarization analysis capability within a ~ 8–10 meV energy window for scattered neutron energies in the thermal range up to ~ 25 meV.

**1. Introduction**

The neutron polarization analysis (PA) [1] is broadly recognized as an important tool in neutron scattering studies. It allows distinguishing between the spin excitations and phonons, separating magnetic scattering from background and uncontrolled structural features ("Braggons", "spurions", etc.), studying magnetic critical dynamics, etc. Recently, PA was successfully applied to the doped strongly correlated oxides, [2,3] quantum magnets, [4,5] studying phase transitions and novel ordered phases in complex systems [6,7].

Traditionally, polarized neutron studies were performed on the crystal (mainly triple axis) spectrometers at steady-state neutron sources and by employing the polarization-dependent Bragg reflectivity of the Heusler-alloy ($Cu_2MnAl$) crystal. The time-of-flight (TOF) spectroscopy in many cases offers significant advantages by surveying neutron scattering events in a wide range of angles and energy transfers in a single measurement. The superiority of the TOF approach is overwhelming at the pulsed (spallation) neutron sources, where the incident neutron beam is inherently time-modulated. However, development of the PA techniques for the TOF instruments is still in its infancy and presents an area of the expected future growth.

The main problem of using the traditional PA techniques in the TOF setup is posed by the contradictory requirements of a large angular acceptance on the one hand, and of the well-collimated beams on the other. Indeed, a beam collimation is required both in the Heusler crystal PA setup [1,8] and in the PA setup with the supermirror-bender transmission polarizers (SBTP) [9]. One venue for the PA on the TOF instruments is opened by the recent progress in developing the polarized $^3$He transmission cells [10]. However, while the great potential of this approach is unquestionable, it has not yet been established as a reliable technique where the large angular apertures are required. An alternative approach consists in using the multi-channel setup by replicating the traditional, collimated-beam PA devices covering the large range of scattering angles. Such scheme is currently employed on the D7 spectrometer at the ILL [10]. Here we describe the multi-channel PA setup with the transmission polarizers proposed for the Hybrid Spectrometer (HYSPEC) at the pulsed Spallation Neutron Source (SNS).

**2. The polarized beam setup on HYSPEC.**

HYSPEC is a direct-geometry, crystal/TOF Hybrid Spectrometer designed for the SNS. It will operate in the thermal and sub-thermal neutron range [2.5, 90] meV, have a resolution comparable to that of a reactor-based triple axis spectrometer, or better, and will have a polarization analysis capability. HYSPEC combines the time-of-flight spectroscopy with the focusing Bragg optics by using the TOF for selecting the neutron energy and the vertically-curved crystal monochromator for concentrating the neutron flux on sample. In this setup, a particular incident neutron polarization needed for the PA can be selected by using the (111) Bragg reflection from a Heusler crystal. This reflection has the property that the nuclear and magnetic scattering lengths are equal so only one spin state is reflected. Studies indicate that the polarization in excess of 95% is achievable when the Mn moments are aligned, and the Bragg reflectivity can approach that expected for an ideal mosaic crystal such as PG [8].

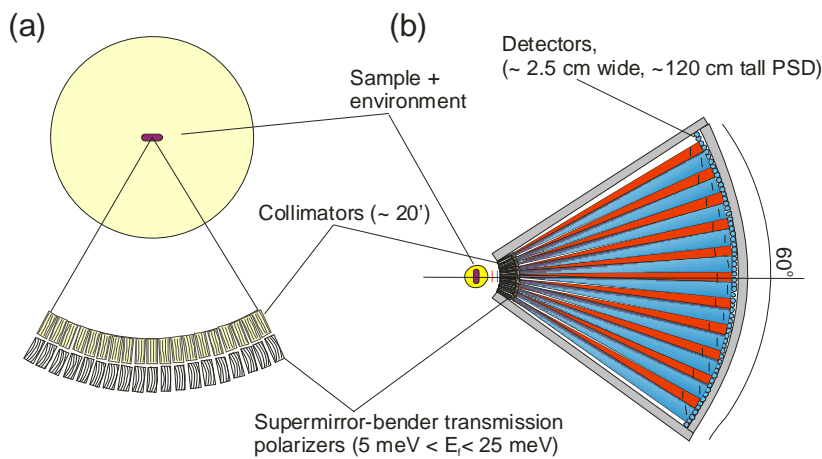

**Figure 1.** Schematics showing (a) geometry of the HYSPEC's multi-channel setup for the scattered beam polarization analysis and (b) the operation of the analyzer in the polarized beam mode. Shading and arrows illustrate how the supermirror benders split scattered neutrons into beams with opposite polarizations.

Large angular acceptance of the HYSPEC's analyzer (a 60º horizontal coverage is currently planned) does not allow using a Heusler crystal for determining the polarization of the scattered neutrons. Therefore, a multi-channel array of equivalent, broadband, supermirror-bender transmission polarizers (SBTP) is envisioned for the polarization analysis of the scattered beam, Figure 1, (a). Nineteen-to-twenty benders could be positioned in front of the analyzer vessel, at a distance ≈ 0.5 m from the sample axis, and within the 60º angle subtended by the detector array. For each PA channel this allows a ≈ 3º sector, or a ≈ 24 cm long segment on the detector bank at 4.5 m from the sample, containing 8-10 detector tubes.

## 3. Performance and optimization of the transmission polarizer.

Supermirror-bender polarization analyzer is a short, multi-channel curved guide with magnetically aligned, polarization-sensitive Fe-Si supermirror films on the channel walls. In practice, each channel is made of a thin, supermirror-coated single-crystal silicon wafer bent to the desired curvature [9]. The neutron critical reflection angle of such (magnetically aligned) film is large, $\theta_c^\uparrow \geq 3.0\, \theta_c^{(Ni)}$, for one spin state and is small, $\theta_c^\downarrow = 0.6\, \theta_c^{(Ni)} \approx 0$, for the other (here $\theta_c^{(Ni)} \approx 0.63°/k_f$ is the critical reflection angle for natural nickel, and $k_f$ is the scattered neutron's wave vector in Å$^{-1}$). Hence, the neutrons of one polarization will follow the curvature of the guide, while those of the other will go essentially straight through. The two polarizations will thus be spatially separated at the detector bank [10,12] and can be measured simultaneously, thereby optimally exploiting the spectrometer's detector coverage, Figure 1, (b). With two SBTP arrays optimized for 10 meV and 20 meV it would be possible to perform the PA of scattered neutrons for energies from ~ 5 meV to ~ 25 meV.

Generic setup of a single transmission polarizer unit is illustrated in Figure 2, (a). It consists of an up-stream collimator and a SBTP. The collimator ensures that neutron beams with different ("up" and "down") polarizations, corresponding to the neutrons "deflected" and "transmitted" by the SBTP (respectively), do not overlap. Hence, the collimation $\eta$ should be smaller than the angular separation between the two beams, $\Delta\theta$, introduced by the polarizer,

$$\eta < \Delta\theta \approx \theta_c^\uparrow - \theta_c^\downarrow, \text{ or, } \eta < 2.4\, \theta_c^{(Ni)} \approx 1.5°/k_f. \tag{1}$$

For a $\eta = 20'$ collimator considered here, Figure 2, (a), this condition is fulfilled for neutron energies below $\approx 41$ meV.

Parameters defining geometry of the individual bender-polarizer are shown in Figure 2. Although there are a fair number of parameters, many of them are coupled and/or constrained. In particular, $\theta_c^\uparrow$ and $\theta_c^\downarrow$ are limited by the available technology, while the length of the SBTP, $L$, must not exceed $\approx 5$ cm if we require that neutron beam attenuation in silicon is less than ≈ 10%. Furthermore, for a given channel length, $L$, its bend angle, $\alpha$, and its curvature radius, $R$, are related through $\alpha = L/R$, and are constrained by the mechanical properties of the silicon wafers and the bending mechanics. Finally, the channel width, $d$, (i. e. the thickness of an individual single-crystal Si wafer) is limited by the requirement of closing the direct line-of-sight through the channel,

$$d \leq R(1 - \cos\alpha) \approx L\alpha/2. \qquad (2)$$

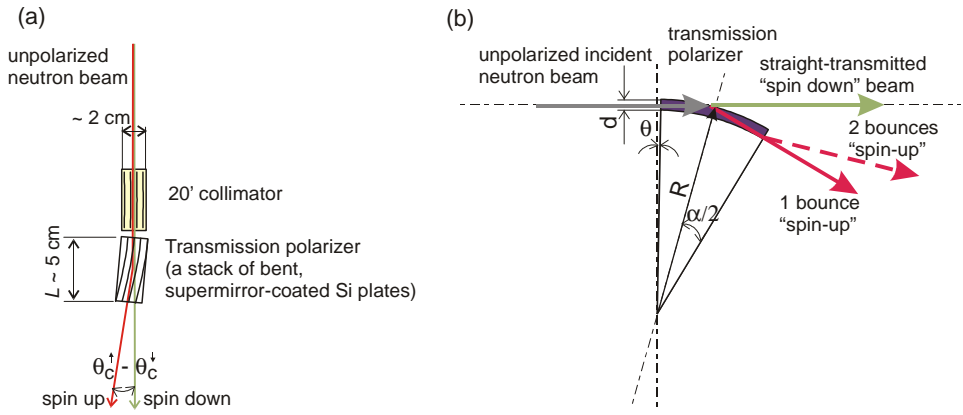

**Figure 2.** (a) Setup of a single PA unit consisting of a collimator and a transmission polarizer. The angular splitting of the two beams is determined by the difference of the critical angles for the corresponding neutron spin polarizations, $\theta_c^\uparrow - \theta_c^\downarrow \geq 2.4\,\theta_c^{(Ni)}$. (b) Geometry of a single channel of the SBTP showing the relevant parameters: the channel width, $d$, the angular size, $\alpha$, the radius of curvature, $R$, and the polarizer tilt angle, $\theta$.

Therefore, the only parameter in the SBTP setup that is free from the technological constraints and can be fully optimized is the polarizer's tilt, or "rocking" angle, $\theta$ [the angle between the polarizer and the collimator's axis, Figure 2, (b)]. Because the SBTP is rather short, it is essentially a single-bounce device (i. e. the most probable neutron passing through the channel is only reflected once). In this case a simple analytical estimate for the tilt angle $\theta$ optimizing the SBTP operation follows from matching the reflection condition for the most probable neutron (i. e. the neutron that travels parallel to the collimator axis) at the channel's end-point,

$$\theta + \alpha = \theta_c^\uparrow \approx 3.0\,\theta_c^{(Ni)}. \qquad (3)$$

Thus, the optimized bender tilt angle is neutron-energy-dependent, and decreases with the increasing energy.

Performance of the SBTP for 15 meV neutrons was experimentally studied by C. Majkrzak in Ref. [9]. He has demonstrated that good polarization sensitivity is achievable if the polarizer tilt angle is appropriately tuned. Here we investigate the performance of the SBTP for different neutron energies using the Monte-Carlo (MC) simulations with NISP package [1]. In view of the above constraints, we adopt the same values of the polarizer bend angle, its length, and the channel width, $L = 5$ cm, $\alpha = 0.57°$ and $d = 0.025$ cm, as used in Ref. [9]. Also, we use $\theta_c^\uparrow = 3.0\theta_c^{(Ni)}$ and $\theta_c^\downarrow = 0.6\theta_c^{(Ni)}$ for the two critical angles.

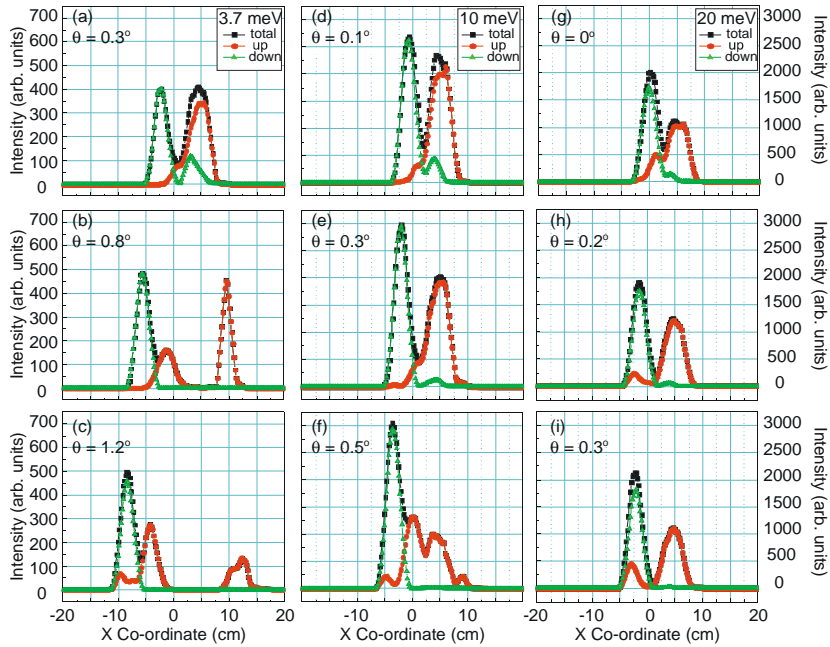

**Figure 3.** Horizontal profile of the neutron intensity at the detector placed 3.9 m behind the polarizer obtained using the Monte-Carlo simulations with NISP package. The „up" (red) and „down" (green) spin polarizations correspond to the deflected and the transmitted neutron beams, respectively. Black symbols show the total intensity. Results for three different values of the polarizer tilt angle θ are shown for three scattered neutron energies, $E_f = 3.7$ meV in (a) – (c), $E_f = 10$ meV in (d) – (f), and $E_f = 20$ meV in (g) – (i). Scale on the right is for the panels (d) – (i).

Our MC results for the horizontal distribution of the transmitted (spin-down, shown in green), deflected (spin-up, shown in red), and total (black) neutron intensities on the HYSPEC's detector bank at ≈ 3.9 m from the polarizer's rear face and for the three neutron energies, 3.7 meV, 10 meV and 20 meV, are shown in Figure 3. It is clear from the figure that the optimized polarizer tilt is smaller at higher energy, where a much finer tuning of this tilt is required. Multiple peaks in the "deflected" channel appearing at higher SBTP tilt angles arise from the consecutive neutron reflections from the channel's walls.

To quantify the PA efficiency of the SBTP setup we have divided the detector in two parts with respect to the minimum of the total neutron intensity which separates the two nearest peaks with opposite polarizations. We then assigned the side containing the straight-transmitted beam to measure the „down" polarization, and the rest of the detector to measure the „up" polarization. The polarization efficiency (PE) was then obtained by dividing the intensity of the selected polarization on each side of the detector by the total neutron intensity on that side.

The resulting SBTP "rocking curves" for the three neutron energies are shown in Figure 4, (a) – (c). For all three energies, the PE of ≈ 90% can be achieved by tuning the SBTP tilt to its optimum value, $\theta_0$. The optimum tilt thus obtained from our MC simulation agrees reasonably well with the simplistic estimate of Eq. (3), which predicts that $\theta_0$ is 0.84°, 0.29°, and 0.04° for neutron energies of 3.7 meV, 10 meV, and 20 meV, respectively. The agreement is better at lower energies, where the neutron critical angles are larger and the PA setup is not very sensitive to the polarizer's alignment. At higher $E$ the critical angle becomes small and Eq. (3) results in an under-illumination of the top SBTP supermirror coating by a divergent neutron beam. In this case the left side of Eq. (3) can be roughly improved by subtracting the half-width of the neutron beam's angular divergence (determined by the collimation), resulting in

$$\theta \approx \theta_c^\uparrow - \alpha + \eta/2. \qquad (4)$$

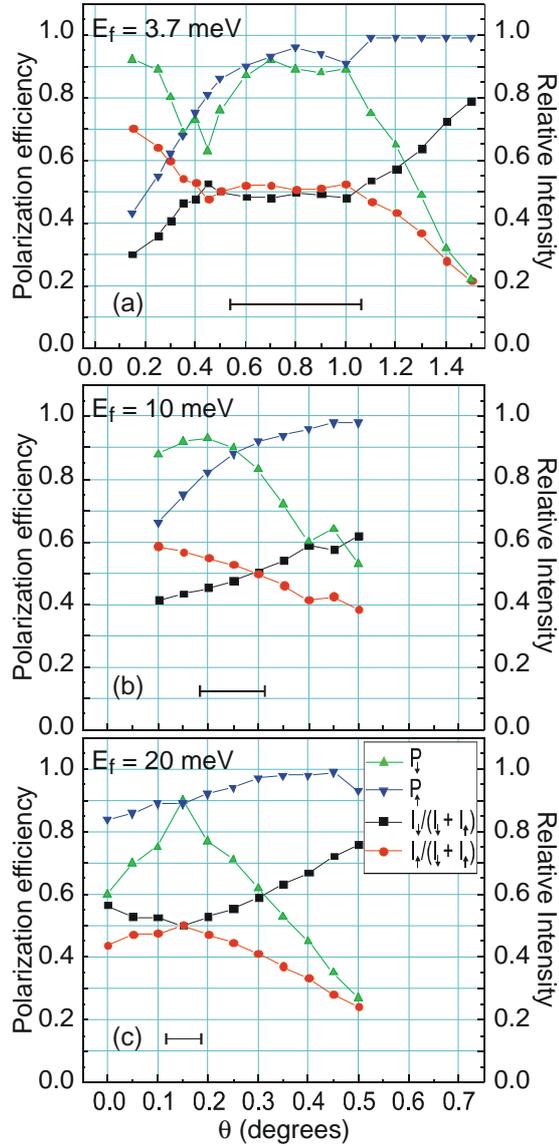

**Figure 4.** The „rocking curves" for the SBTP obtained using the NISP MC simulations similar to those shown in Figure 3. For each $E_f$ and θ we used the minimum in the total neutron intensity (black symbols in Figure 3) which separates the two nearest peaks with different polarizations to divide the detector in two parts that measure the respective polarizations. The polarization efficiency (PE) was then obtained by dividing the intensity of the selected polarization on each side of the detector by the total neutron intensity on that side. The relative intensity shows the relative contribution of „up" and „down" polarizations over the entire detector. The horizontal bars show the angular range where PE > 80% [note the different x-axis scales in the panel (a) and the panels (b), (c)].

With the increasing SBTP tilt the PE of the deflected beam grows, but its relative intensity decreases. The angular region where the PE is > 80% is shown by the horizontal bars in

Figure 4, (a) – (c). For the lower neutron energy, $E = 3.7$ meV, this "working" region is quite large, indicating that the setup is rather un-sensitive to the SBTP alignment. However, this region shrinks rapidly with the increasing neutron energy. At higher energies, above $\approx 10$ meV, the PA setup proposed here would require a fine tuning of the polarizer's rocking angle.

**4. Summary.**

We have described the polarized beam setup proposed for the Hybrid Spectrometer at the SNS. In this setup the polarization analysis of the scattered neutrons is carried out by a multi-channel array of the supermirror-bender transmission polarizers. We have studied the performance of such a polarizer for different neutron energies using the Monte-Carlo ray-tracing simulations. Our results show that the polarization efficiency of up to 90% is achievable for neutron energies at least up to 20 meV with appropriate tuning of the SBTP rotation angle with respect to the up-stream collimator. Furthermore, an acceptable PE > 80% is achievable within a rather broad energy window, $\Delta E \sim 8 – 10$ meV, when this angle is appropriately aligned, optimizing the SBTP performance for a particular neutron energy within this window.

**Acknowledgements**

We thank S.-H. Lee, T. Krist, and M. Hagen, for useful remarks and discussions. This work was supported by the US DOE under the Contract DE-AC02-98CH10886.